\documentclass[11pt, letterpaper]{elsarticle}
\usepackage[utf8]{inputenc}
\usepackage[margin=1.5cm]{geometry}
\usepackage{titlesec}
\usepackage{tabu}
\usepackage{enumitem}
\usepackage{amssymb}
\usepackage{xcolor}
\usepackage{paralist}
\usepackage{microtype}
\usepackage[numbers]{natbib}
\newlist{selectlist}{itemize}{2}
\setlist[selectlist]{label=$\square$,leftmargin=*,noitemsep,topsep=0pt}

\usepackage[normalem]{ulem}
\renewcommand{\sout}[1]{}
\newcommand{\hl}[1]{\textcolor{black}{#1}}

\usepackage{lmodern}

\usepackage{hyperref}
\hypersetup{
    colorlinks=true,
    linkcolor=blue,
    filecolor=magenta,      
    urlcolor=blue,
}
 
\titleformat{\section}[block]{\vskip0.5cm\bfseries}{\thesection.}{0.5em}{} 
\titleformat{\subsection}[block]{}{\thesubsection}{0.5em}{}

\usepackage{lineno}

\journal{Software Impacts}

\begin{document}

\noindent
\textbf{\textit{libACA, pyACA, and ACA-Code: Audio Content Analysis in 3 Languages}}
\vskip0.5cm
\noindent
\textbf{\textit{Alexander Lerch}}\\

\section*{Abstract}
The three packages libACA, pyACA, and ACA-Code provide reference implementations for basic approaches and algorithms for the analysis of musical audio signals in three different languages: C++, Python, and Matlab. All three packages cover the same algorithms, such as extraction of low level audio features, fundamental frequency estimation, as well as simple approaches to chord recognition, musical key detection, and onset detection. In addition, it implementations of more generic algorithms useful in audio content analysis such as dynamic time warping and the Viterbi algorithm are provided. The three packages thus provide a practical cross-language and cross-platform reference to students and engineers implementing audio analysis algorithms and enable implementation-focused learning of algorithms for audio content analysis and music information retrieval.

\section*{Keywords}
Audio Content Analysis, Music Information Retrieval, C++, Python, Matlab
\newpage
\section*{Code metadata}
\begin{tabular}{|l|p{6.5cm}|p{9.5cm}|}
\hline
\textbf{Nr.} & \textbf{Code metadata description} & \textbf{Please fill in this column} \\
\hline
C1 & Current code version & v0.3.1 \\
\hline
C2 & Permanent link to code/repository used for this code version & \href{https://github.com/alexanderlerch/2022-ACA-SIMPAC}{github.com/alexanderlerch/2022-ACA-SIMPAC}\\
\hline
C3  & Permanent link to Reproducible Capsule & \href{https://codeocean.com/capsule/b1eb9884-ac26-4292-b752-f7e02280f38c/}{codeocean.com/capsule/b1eb9884-ac26-4292-b752-f7e02280f38c}\\
\hline
C4 & Legal Code License   & MIT License \\
\hline
C5 & Code versioning system used & git \\
\hline
C6 & Software code languages, tools, and services used & C++, Python, Matlab\\
\hline
C7 & Compilation requirements, operating environments \& dependencies & \begin{footnotesize}\vspace{1mm}\shortstack[l]{C++: CMake with Gcc, Clang, or MSVC, C++11\\ Python: Python 3.x, Dependencies: numpy, scipy, matplotlib\\ Matlab: V2016x+, Dependencies: Signal Processing Toolbox}\end{footnotesize}\\
\hline
C8 & If available Link to developer documentation/manual & \shortstack[l]{  \href{https://alexanderlerch.github.io/libACA}{alexanderlerch.github.io/libACA}\\ 
                                                                            \href{https://alexanderlerch.github.io/pyACA/}{alexanderlerch.github.io/pyACA/}\\ 
                                                                            \href{https://alexanderlerch.github.io/ACA-Code/}{alexanderlerch.github.io/ACA-Code/}}\\
\hline
C9 & Support email for questions & \href{mailto:info@audiocontentanalysis.org}{info@audiocontentanalysis.org}\\
\hline
\end{tabular}\\

\section{Introduction}
The areas of audio content analysis and music information retrieval are growing fields with evolving demands on software solutions available to researchers, teachers, software engineers, and system designers. There are, therefore, different objectives that a software can aim to address. Such objectives include, for example, to efficiently solve a well-established problem, to design state-of-the art systems for potentially new tasks, to use the software package in educational environments to demonstrate algorithmic properties, or to provide reference implementations of established algorithms. Often, these objectives are not easily aligned, i.e., a software for problem solving might not be useful for classroom teaching. \sout{Of the existing open source frameworks for audio content analysis, software that targets simple execution for various practical tasks such as libROSA \citep{mcfee_librosa_2015} or Essentia \citep{bogdanov_essentia_2013} is widely used. An example for software targeting  classroom teaching are M\"uller and Zalkow's FMP notebooks \citep{muller_libfmp_2021}.}

\hl{The software presented here implements many tasks related to audio analysis with a focus on educational aspects. It provides reference implementations of multiple audio analysis algorithms such as feature extraction, fundamental frequency detection, chord detection, etc.; what makes the software truly unique is that these reference implementations are available in three different programming languages: C++, Python, and Matlab. Closely integrated with other educational materials such as lecture slides\footnote{\href{https://github.com/alexanderlerch/ACA-Slides/tree/2nd_edition}{github.com/alexanderlerch/ACA-Slides/tree/2nd\_edition/}, last accessed: Jun 28, 2022} and based}
 on the algorithmic introductions in the text book ``An Introduction to Audio Content Analysis'' \cite{lerch_introduction_2012}, the three presented packages \textit{libACA} \citep{lerch_libaca_2022}, \textit{pyACA} \citep{lerch_pyaca_2022}, and \textit{ACA-Code} \citep{lerch_aca-code_2022} aim at giving students and engineers quick access to algorithmic implementations, fostering the understanding of practical implementation details that might otherwise be missed. Thus, the presented three software packages are not in competition \hl{with popular existing open source frameworks for audio content analysis such as libROSA \citep{mcfee_librosa_2015} or Essentia \citep{bogdanov_essentia_2013} which target the easy utilization for audio analysis, but complement these} existing packages by providing easy access to algorithms through implementation in three languages. This allows for a better separation of language-specific choices from algorithmic details. \hl{It also provides an implementation focus that differentiates the ACA packages from classroom software such as M\"uller and Zalkow's FMP notebooks \citep{muller_libfmp_2021}.}

\sout{The three programming languages all have advantages and disadvantages. Matlab/Octave has been traditionally very prominent in the signal processing community and provides a monolithic environment well-suited to quick prototyping and visualization. Python, probably the most widely used language in the field now, is appealing because it combines easy extensibility through modular packages, including easy availability of powerful machine learning packages. C++ is, while not particularly suitable for algorithmic prototyping and lacking easy extensibility, a language that is often used in production environments with high performance requirements and is probably the most commonly used language in audio software. Having implementations in all three languages increases versatility and accessibility and fosters better algorithmic understanding.}

\section{Software description}

\subsection{Functionality}
The functionality of all three packages, whether Matlab, Python, or C++, is identical. The software offers reference implementations for common low level spectral audio features such as Spectral Centroid, Crest Factor, Flatness, MFCCs, Pitch Chroma, Rolloff, and Skewness as well as time-domain features such as RMS and Zero-Crossing-Rate. It also implements multiple fundamental frequency extraction algorithms based on auto correlation, the harmonic product spectrum, and others. Example implementations of chord and key detection, audio fingerprint extraction, beat histogram computation, and onset detection, are complemented by reference implementations of more general algorithms such as dynamic time warping, the Viterbi algorithm, a K nearest neighbor classifier, a Gaussian mixture model,  K-means clustering, and principal component analysis. Sophisticated machine learning methods are not utilized to avoid third-party dependencies (see below).
\hl{One of the core properties of the presented software is that all algorithms are implemented in three different programming languages C++, Python, and Matlab, each of which has specific advantages and drawbacks.
 Matlab/Octave has been traditionally very prominent in the signal processing community and provides a monolithic environment well-suited to quick prototyping and visualization. 
Python, probably the most widely used language in the field now, is appealing because it combines easy extensibility through modular packages, including easy availability of powerful machine learning packages. 
C++ is, while not particularly suitable for algorithmic prototyping and lacking easy extensibility, a language that is often used in production environments with high performance requirements and is probably the most commonly used language in audio software. 
Having implementations in all three languages increases versatility and accessibility and fosters better algorithmic understanding.}

\subsection{Design choices}
The source code aims at demonstrating how to implement audio analysis solutions and at facilitating algorithmic understanding. Thus, the design principles need to be somewhat different from software to be used in production environments. Most importantly, performance optimization is given a very low priority. High priority is assigned to
\begin{inparaenum}[(i)]
	\item   readability, i.e., easily understandable and consistent variable, function, and file naming,
    \item   implementing source code that can be executed on a variety of operating systems and environments (C++ 11 compatibility, Python 3.x compatibility, Matlab 2016 compatibility), and
    \item   reducing third-party dependencies to avoid code/algorithm obfuscation, allow for easier maintainability, and ensure a code base with consistent programming style.
\end{inparaenum}
Note, however, that avoiding third-party dependencies tends to increase the necessary number of lines of code ---~especially in the case of the C++ package~--- and shifts the focus towards \hl{baseline} approaches with low and medium level of sophistication. Deep learning approaches, for instance, are generally avoided.

\subsection{Future extensions}
In the short-term, the packages will be extended with examples showcasing existing functionality in various tasks. For instance, a simple music style classifier can be built with the features, and one of the implemented classifiers, and audio alignment can be implemented utilizing the dynamic time warping code. Simple extensions of functionality can build, for example, on the non-negative matrix factorization to implement multi-pitch detection or drum transcription systems. The highest priority for new implementations from scratch is the currently missing task of beat tracking, a core task in music information retrieval.

\section{Software impact}
\hl{Audio content analysis includes a multitude of research areas, including:
    \begin{inparaenum}[(i)]
        \item   \textit{speech analysis}, e.g., automatic speech recognition \cite{huang_historical_2014} or recognizing emotion in speech \cite{el_ayadi_survey_2011},
        \item   \textit{urban sound analysis}, e.g., noise pollution monitoring \cite{bello_sound_2018} or the automatic detection of dangerous events \cite{crocco_audio_2016},
        \item   \textit{industrial sound analysis}, e.g., monitoring the state of mechanical devices like engines \cite{grollmisch_sounding_2019} or monitoring the health of livestock \cite{berckmans_animal_2015}, and
        \item   \textit{musical audio analysis}, targeting the understanding and extraction of musical parameters and properties from the audio signal \cite{ellis_extracting_2006}.
    \end{inparaenum}
    The presented software can be applied to all of these scenarios and can serve as an educational resource as well as a foundation for research and commercial applications in these areas. It has been used in institutions for higher education.}
Despite the focus on educational aspects, the software can be and has been used in a wide variety of research scenarios, ranging from singing transcription \cite{yang_singing_2015} and crowd behavior estimation \cite{butler_classifying_2018} over EEG signal processing \cite{vinay_mind_2021} to the analysis of sound imitation capacities of orca whales \cite{abramson_imitation_2018}.

\section{Conclusion}
While implementations of most audio analysis algorithms can be found online, their usefulness for understanding algorithmic details is often hampered by different coding styles, naming conventions, extended or removed functionality, and the excessive use of third-party dependencies. The packages \textit{libACA}, \textit{pyACA}, and \textit{ACA-Code} address this problem by showcasing the implementation of a multitude of algorithms for audio content analysis in a consistent way with identical functionality in three different languages: C++, Python, and Matlab. 

\section{Acknowledgments}
The author would like to thank Kaushal Sali (github: \href{https://github.com/kaushalsali}{@kaushalsali}) for his contributions to both \textit{pyACA} and \textit{ACA-Code}.

\section{References}
\renewcommand{\bibsection}{}

\bibliographystyle{IEEEtranN}
\bibliography{2022-libACA}

\end{document}